\documentclass[
    ,final            
  ]
  {aipproc}

\layoutstyle{6x9}

\newcommand{\ie}{{\em i.e.}}

\newcommand{\ccbar}{$c\bar{c}$}

\newcommand{\pythia}{{\sc Pythia}}

\newcommand{\xf}{$x_{F} \,$}
\newcommand{\pt}{$p_{\perp} \,$}

\newcommand{\etal}{{\it et al.}}
\newcommand{\jpsi}{$J/\psi$}
\newcommand{\psip}{${\psi}^{\prime}$}

\begin{document}

\title{Soft and hard QCD in charmonium production
\thanks {Based on the talk
presented by C.~Brenner~Mariotto at the Pan American Advanced Studies Institute
(PASI 2002), Campos do Jord\~ao, Brazil, January 7-18, 2002.}
}

\author{{{C.~Brenner~Mariotto}}}{
  address={Inst.~of~Physics,~Univ.~Fed.~do~Rio~Grande~do~Sul,~Box~15051,~CEP~91501-960~Porto~Alegre,~Brazil}
}

\author{M.B. Gay Ducati}{
  address={Inst.~of~Physics,~Univ.~Fed.~do~Rio~Grande~do~Sul,~Box~15051,~CEP~91501-960~Porto~Alegre,~Brazil}
}

\author{G.~Ingelman}{
  address={High Energy Physics, Uppsala University, Uppsala, Sweden and 
  DESY, Hamburg,~Germany}
}

\begin{abstract}
Hard and soft QCD dynamics are both important in charmonium hadroproduction, as presented here through a next-to-leading order QCD matrix element calculation combined with the colour evaporation model. Observed $x_F$ and $p_\perp$ 
distributions of $J/\psi$ in hadroproduction are reproduced. 
Quite similar results can also be obtained with a Monte Carlo 
event generator where \ccbar \ pairs are instead produced through leading order matrix elements and the parton shower approximation of higher order processes. The soft dynamics may alternatively be described by the soft colour interaction model. We also discuss the relative rates of different charmonium states and introduce an improved model for mapping the continuous \ccbar \ mass spectrum on the physical charmonium resonances. 
\end{abstract}

\maketitle

\vspace{-.3cm}
The theoretical description of charmonium production separates the hard and soft parts of the process based on the factorisation theorem in QCD. Thus, we first consider the perturbative production of a \ccbar \ pair at the parton level and  then the non-perturbative formation of a bound charmonium state \cite{epjc}. 

Perturbative QCD (pQCD) should be applicable for \ccbar \ production, since the charm quark mass $m_c$ is large enough to make $\alpha_s(m_c^2)$ a small expansion parameter. The leading order (LO) processes are $gg \rightarrow c\bar{c}$ and $q\bar{q} \rightarrow c\bar{c}$. The next-to-leading order (NLO) processes, \ie \ ${\cal O}(\alpha_s^3)$, include the emission of a third parton and virtual corrections (where divergences are properly cancelled). The full NLO matrix elements, with explicit charm quark mass, are available in a computer program \cite{NLO} giving total and differential cross sections. 

An alternative description of the pQCD production of \ccbar \ pairs is given by the  \pythia \ \cite{Pythia} Monte Carlo, where all LO QCD $2\to 2$ processes are included with their corresponding matrix elements and the incoming and outgoing partons may branch as described by the DGLAP equations. A \ccbar \ pair can then be produced as described by the LO matrix elements for $q\bar{q} \rightarrow c\bar{c}$ and $gg \rightarrow c\bar{c}$ (with explicit $m_c$ dependence) or in a gluon splitting $g \rightarrow c\bar{c}$ in the parton shower. 

The main free parameter is the charm quark mass $m_c$, taken as $m_c=1.5$~GeV in the NLO program and $m_c=1.35$~GeV in \pythia.  In both approaches, the factorization and renormalization scales are taken as the average transverse mass of the $c$ and $\bar{c}$. 

The formation of bound hadron states occurs through processes with small momentum transfers such that $\alpha_s$ is large and prevents the use of perturbation theory. The lack of an appropriate method to calculate non-perturbative processes, forces us to use phenomenological models to describe the formation of charmonium states from perturbatively produced \ccbar \ pairs. The Color Evaporation Model (CEM) \cite{HALZENquantit} and the Soft Colour Interaction (SCI) model \cite{sci,sci-onium} are based on a similar phenomenological approach, where soft colour interactions can change the colour state of a \ccbar \ pair from an octet to a singlet. They employ the same hard pQCD processes to produce a \ccbar \ pair regardless of its spin state. A colour singlet \ccbar \ pair with an invariant mass below the threshold for open charm ($m_{c\bar{c}}<2m_D$) will then form a charmonium state.

In CEM \cite{HALZENquantit,satz,schuler,CEMnosso} the exchange of soft gluons is assumed to give a randomisation of the colour state. This implies a probability $1/9$ that a $c\bar{c}$ pair is in a colour singlet state and produces charmonium if its mass is below the threshold for open charm production, $m_{c\bar{c}}<2m_D$. The fraction of a specific charmonium state $i$, relative to all charmonia, is given by a non-perturbative parameter $\rho_{i}$ ($\rho_{J/\psi}=0.4-0.5$) \cite{HALZENquantit}.

In SCI \cite{sci,sci-onium,gaps} it is assumed that colour-anticolour, corresponding to non-perturbative gluons, can be exchanged between partons emerging from a hard scattering and hadron remnants. The unknown probability to exchange a soft gluon between parton pairs is given by a phenomenological parameter $R$. These colour exchanges lead to different topologies of the confining colour string-fields and thereby to different hadronic final states after hadronisation. The mapping of $c\bar{c}$ pairs, with mass below the threshold for open charm production, is here made based on spin statistics  resulting in a fraction of a specific quarkonium state $i$ with total angular momentum $J_i$ given by $f_i = \frac{\Gamma_i}{\sum_k \Gamma_k} $, where $\Gamma = (2J_i+1)/n_i$ including a suppression of radially excited states through the main quantum number $n_i$. This model was found to give a correct description of the different heavy quarkonium states observed at the Tevatron \cite{sci-onium}. 

The complete models are formed by adding the CEM or SCI models for the soft processes to any of the descriptions for the hard pQCD processes. The first model we label {\bf CEM-NLO} and is the combination of the CEM model with the NLO program. The second model is {\bf CEM-PYTHIA}, where CEM has been implemented in \pythia \ version 5.7 \cite{Pythia}. The third model, {\bf SCI-PYTHIA}, is to use the SCI model as implemented in \pythia \ 5.7.  Further ingredients are the intrinsic $k_{\perp}$, due to the Fermi motion of partons inside the initial state hadrons, and soft $p_T$ in soft gluon exchange that neutralize color. Both effects are modelled by a gaussian distribution of width $0.6 - 0.8$~GeV used in \pythia \ and in the NLO program.

Comparing these three models we can separate different effects. With CEM implemented in the NLO program and in \pythia, we can compare the pQCD  contributions, namely NLO versus LO plus the parton shower approximation of higher orders. Having SCI and CEM implemented in \pythia, we can explicitly compare these two non-perturbative models and see to what extent they can account for observed soft effects.

Detailed comparisons between the models have been done as well as extensive comparison with data, both from fixed target experiments and the Tevatron collider \cite{epjc}. Here we limit ourselves to proton beams. The targets are different nuclei, but the experimental results are rescaled to the cross section per nucleon. Thus we compare directly with our models which do not include any nuclear effects but treat hadron-nucleon interactions. 

\begin{figure}[ht]
  \includegraphics[height=.625\textheight]{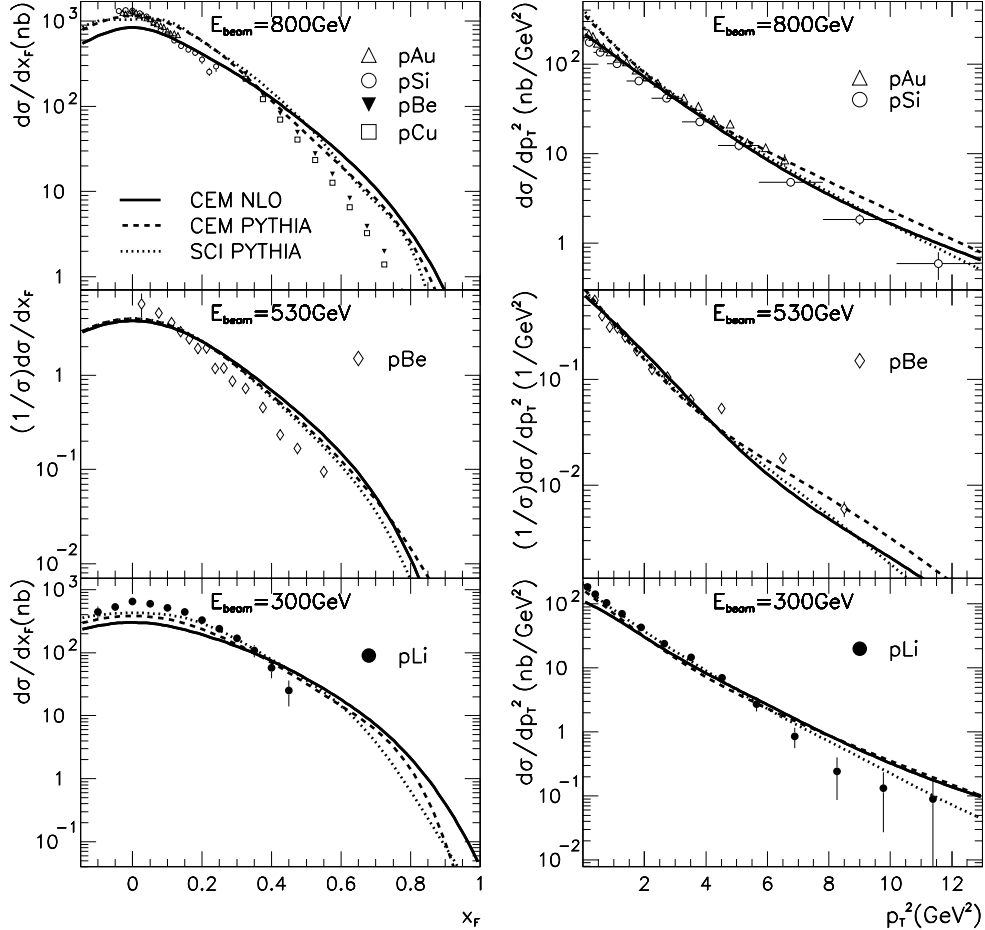}
\caption{Distributions in $x_F$ and $p_\perp^2$ of $J/\psi$ produced with proton beams of energies $800$, $530$ and $300\, GeV$ on fixed target. Data \protect \cite{pp800,300,pp800k,pp530} compared to CEM based on NLO pQCD matrix elements, and CEM and SCI based on LO matrix elements plus parton showers in the \pythia \ Monte Carlo.\label{pp}}
\end{figure}

Fig.~\ref{pp} shows $x_F$ and $p_\perp$ distributions of the produced $J/\psi$ for proton beams of different energies. As can be seen, the data are approximately reproduced, both in shape and normalization, by all three models. Looking into the details of the \xf \ distributions, one can observe that the model curves fall less steeply than the data and therefore overshoot somewhat at large \xf. The observed \pt distribution is better reproduced, with only small differences between the models. 

\begin{figure}[ht] 
\begin{tabular}{c c}
  \includegraphics[height=.263\textheight]{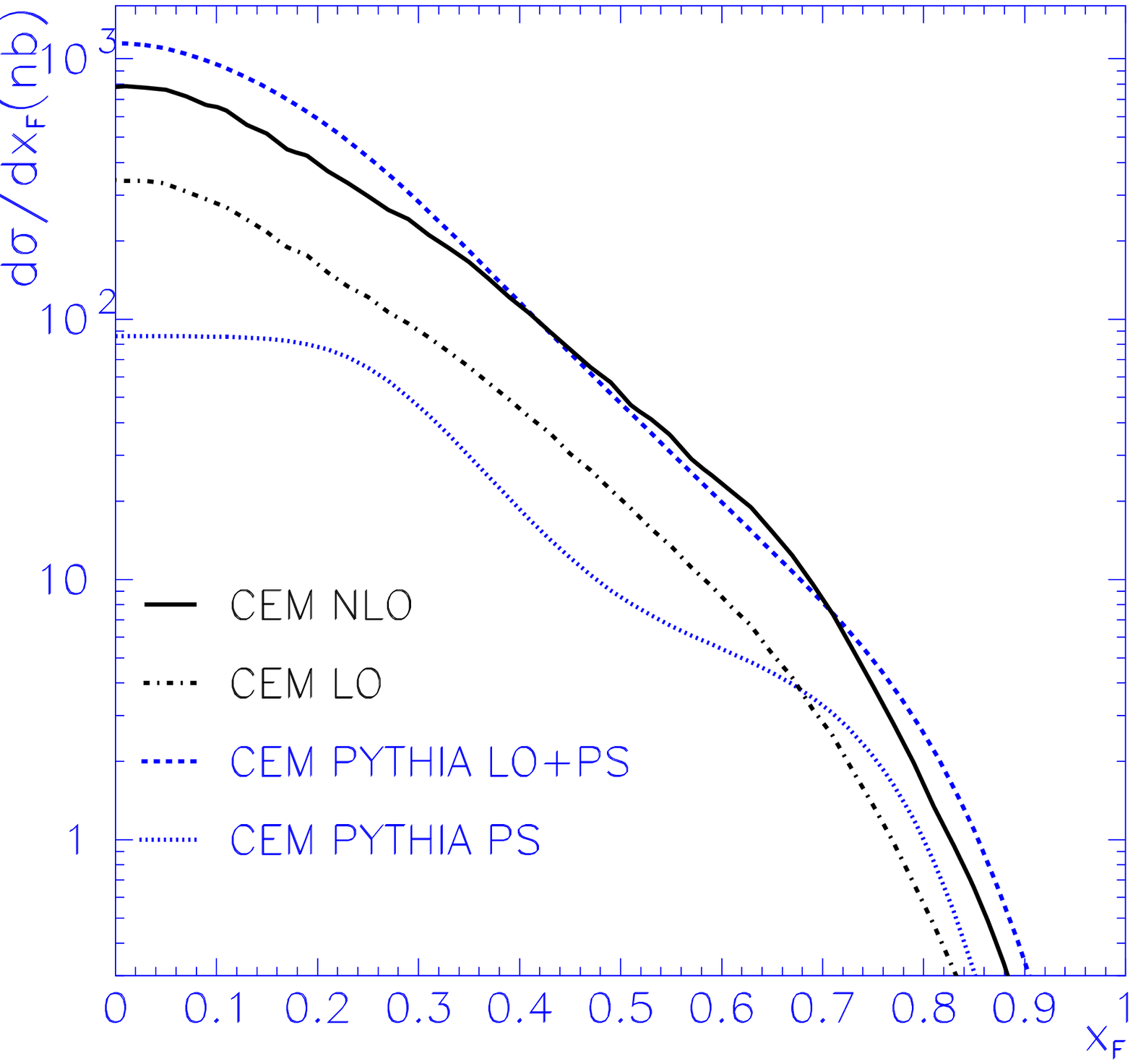}
  \includegraphics[height=.263\textheight]{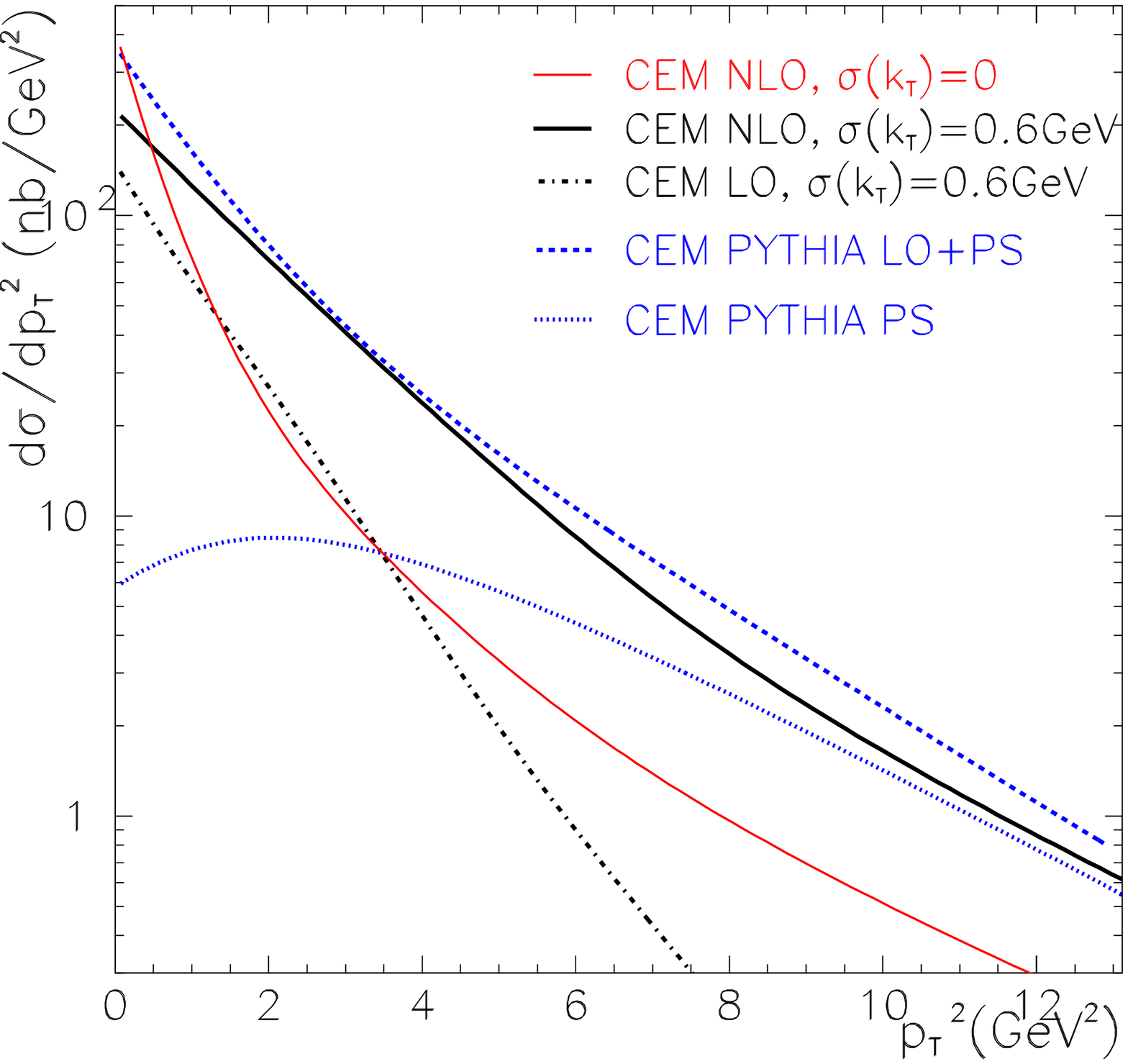}
\end{tabular}\vspace{-0.5cm}
\caption{Distributions in $x_F$ and $p_\perp^2$ of $J/\psi$ (in 800~GeV proton on proton as in Fig.~\ref{pp}) from variations of the pQCD treatment.\label{fig:xfteori} CEM based on the NLO program with $m_c=1.5$~GeV: NLO and LO matrix elements, NLO with no intrinsic $k_\perp$. CEM based on \pythia \ with $m_c=1.35$~GeV: LO matrix elements plus parton showers (PS) and PS contribution shown separately.}
\vspace{-.5cm}
\end{figure} 

Having CEM combined with different treatments of the pQCD production of \ccbar , we can now investigate pQCD effects in more detail. Fig.~\ref{fig:xfteori} illustrates this for the case of 800~GeV proton energy, similar conclusions can also be drawn for other energies and beam particles.
For the \xf \ distribution in Fig.~\ref{fig:xfteori}a, the full NLO result and that based on LO+PS agree reasonably well.
The NLO corrections are very important, as we see by comparing the LO and the full NLO results. In the LO+PS result, however, the PS contribution is unimportant for the overall cross section which is dominated by the LO \ccbar \ production. The agreement with the NLO result is here obtained by using a lower charm mass, $m_c=1.35$~GeV. We have cross-checked this within the NLO program, 
where the full result is essentially reproduced by the LO part if this lower mass value is used. This demonstrates that the NLO correction is essentially an overall $K$-factor from soft and virtual corrections.
For the \pt distributions in Fig.~\ref{fig:xfteori}b, the NLO program gives a \pt distribution with a much larger tail at large \pt, but it is still substantially affected by the inclusion of the intrinsic $k_\perp$ at the limited values of \pt accessible at fixed target energies. The \pt distribution resulting from the LO+PS in the \pythia \ approach, is at high-\pt dominated by 
\ccbar \ from gluon splittings in the partons showers, whereas the bulk of the cross section comes from the low-\pt region where the LO diagrams dominate. The total LO+PS result, which also includes a gaussian intrinsic $k_\perp$ with the same width 0.6~GeV, agrees quite well with the NLO result.

Data on \psip \ production provide an additional testing ground for the models, which produce all charmonium states with the same dynamics. A comparison made in \cite{epjc} shows that all models account quite well for the shape of the distributions. The proper normalization of CEM is obtained by chosing  $\rho_{\psi^{\prime}}=0.066$. The spin statistics used in SCI predicts only a factor two suppression of \psip , and must be lowered by an additional factor four in order to reproduce the data. This has prompted us to develop a more elaborate model for turning \ccbar \ pairs into different charmonium resonances \cite{epjc}, which is briefly described here.

\begin{figure}[t]
  \includegraphics[height=.3\textheight]{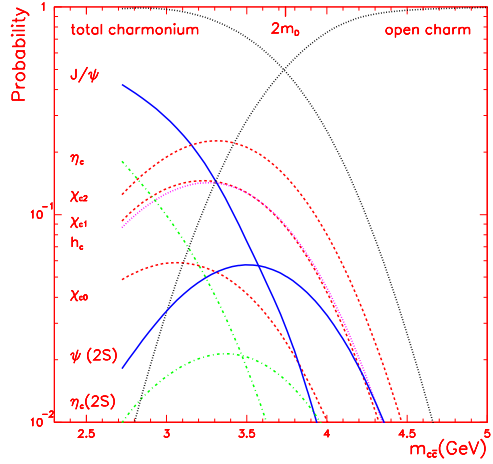}
  \caption{Probability distributions for the different charmonium states as obtained in the model with gaussian smearing ($\sigma_{sme}=400\, MeV$). The resulting total probability for charmonium production and the remainder as open charm production are also shown.}
\label{dQ2PsDserr}
\vspace{-0.5cm}
\end{figure}

The \ccbar \ pair is produced in a pQCD process with a continuous distribution of its invariant mass $m_{c\bar{c}}$ and must be mapped onto the discrete spectrum of charmonium states. The soft interactions that turn the pair into a colour singlet and form the state, may very well change its mass by a few hundred MeV, which is the typical scale of the soft interactions. We model this by a gaussian smearing of a few hundred MeV. The probability to end up in a specific resonance, shown in Fig.~\ref{dQ2PsDserr}, is then proportional to the superposition of this gaussian with the resonance peak, times the corresponding spin-statistics factor. The smearing of $m_{c\bar{c}}$ across the threshold $2m_D$ for open charm, implies non-zero contributions for charmonium also above the $D\overline{D}$ threshold as well as some open charm production for $m_{c\bar{c}}$ originally below this threshold. 

\begin{figure}[t]
\begin{tabular}{c c}
  \includegraphics[height=.30201\textheight]{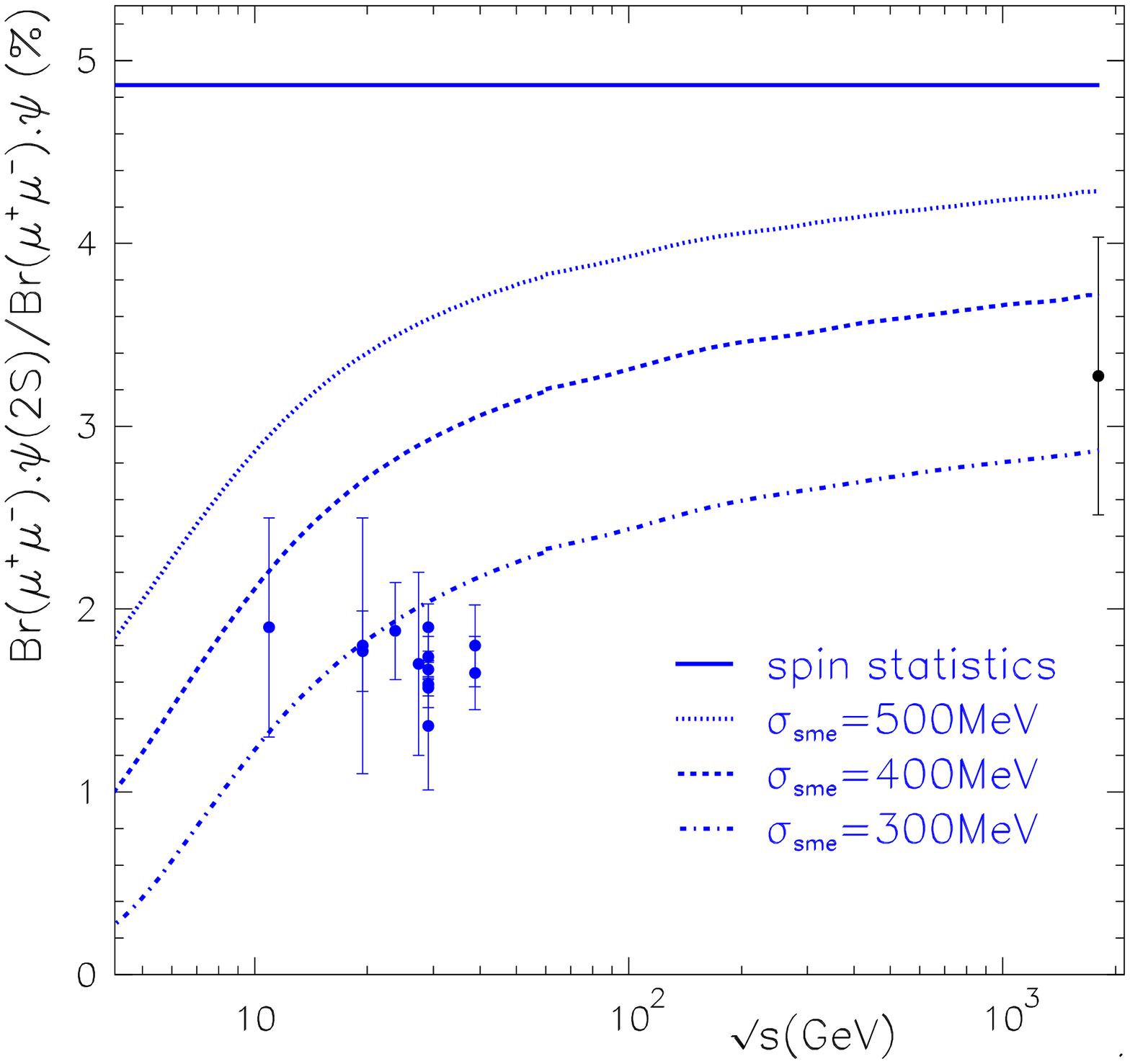}
  \includegraphics[height=.39401\textheight]{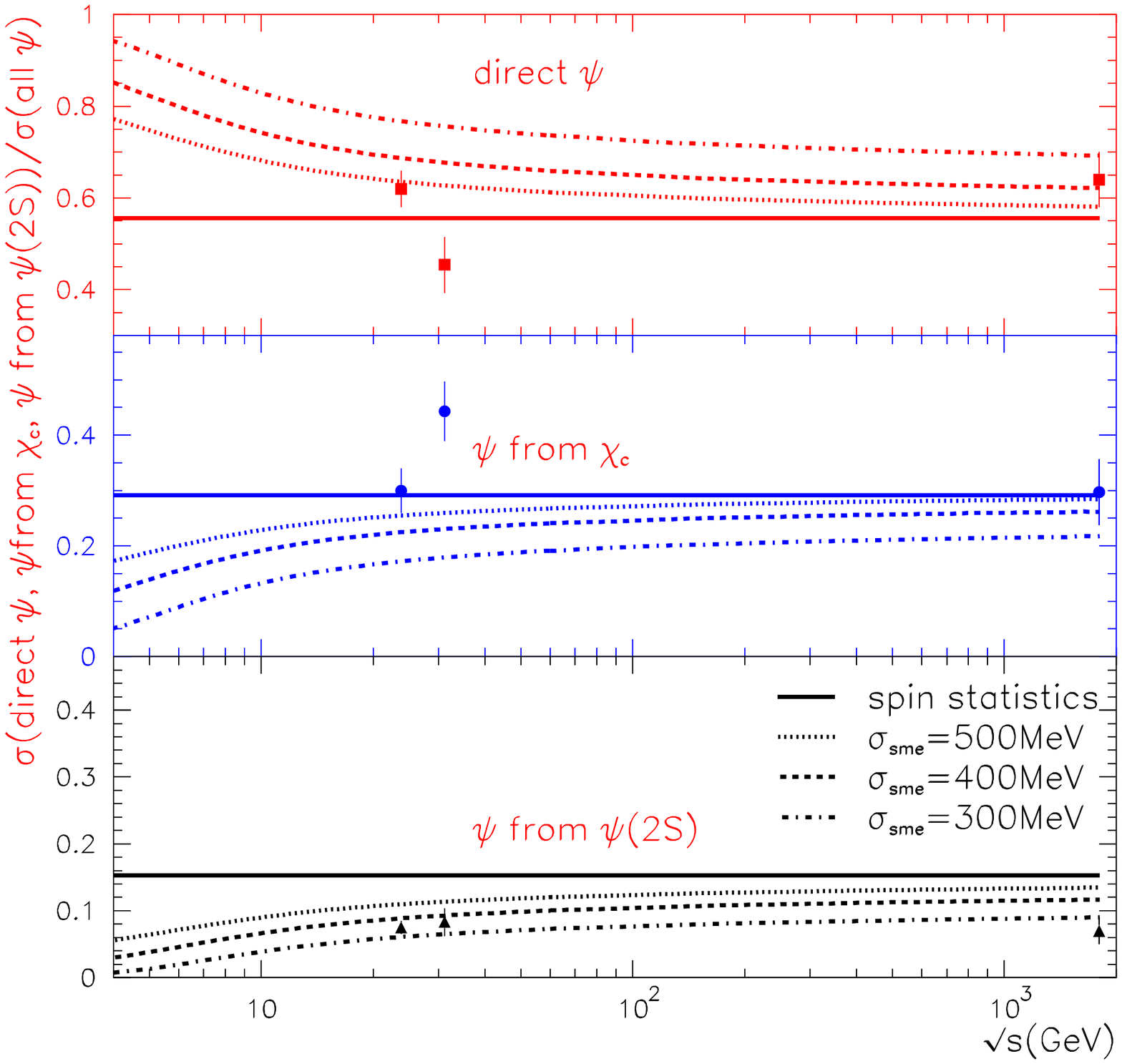}
\end{tabular}\vspace{-0.5cm}
\caption{The ratio of \psip \ to \jpsi \ production (times their branching ratios for decay into $\mu^+\mu^-$) (left) and fractions of \jpsi \ produced directly, and coming from the decay of $\chi_c$ and $\psi^{\prime}$ states (right) in hadron-hadron interactions of cms energy $\sqrt{s}$. Data \cite{Na38p450,high-pt-data,ppi300E705, pi515koreshev} compared to simple spin statistics and to our model with different gaussian smearing widths applied to CEM.}
\label{dQ2gaussfract}
\end{figure}

By folding these charmonium probability functions with the distribution in $m_{c\bar{c}}$ obtained from pQCD, one gets the cross section for a given charmonium state. Applying this mapping procedure to the CEM model we obtain the results in Fig.~\ref{dQ2gaussfract}. As opposed to the simple spin statistics factor, this model gives a reasonable description of the observed ratio of \psip \ to \jpsi \ production and fractions of \jpsi \ produced directly, coming from decays of $\chi_c$ states and from $\psi^{\prime}$. In particular, the model gives a characteristic energy dependence of the kind indicated by the data. 

In summary, both hard and soft QCD dynamics play important roles in the production of charmonium states in hadronic interactions. The \ccbar \ pair production in pQCD have substantial higher order contributions, with a factor two increase of the total cross section from NLO corrections. These come mainly from soft and collinear gluon emissions combined with virtual corrections and can be effectively accounted for by an overall $K$-factor. This supports to the use of the \pythia \ Monte Carlo with LO matrix elements and a reduced charm quark mass to increase the cross section correspondingly. The high \pt tail of the cross section is, however, dominated by higher order tree diagrams in the NLO matrix elements and in the parton showers of the Monte Carlo approach. 

The non-perturbative formation of \jpsi , can be described by the Colour Evaporation Model and the Soft Colour Interaction model, where \ccbar \ pairs in a colour octet state can turn into a colour singlet state by soft gluon exchange. A simple spin statistics factor is not sufficient for a proper description of other charmonium states, but our more elaborated model to map \ccbar \ pairs onto the physical charmonium states improves this situation. 

To conclude, the main features of hadroproduction of charmonium can be described in these models combining pQCD and effects of soft colour exchanges. This shows, in particular, that these models for the soft QCD dynamics contain the essential effects and therefore improve our understanding of non-perturbative QCD. 

\vspace{-.4cm}
This work was partially financed by Funda\c{c}\~ao Coordena\c{c}\~ao de Aperfei\c{c}oamento de Pessoal de N\'{\i}vel Superior (CAPES), Brazil, and by the Swedish Research Council.

\vspace{-.6cm}
\bibliographystyle{aipprocl} 

\end{document}